# Signal, Noise, and Burnout: A Human-Information Interaction Analysis of Voter Verification in a High-Volatility Environment

Kijung Lee

**Abstract**
The 2024 U.S. Presidential Election unfolded within an information environment of unprecedented volatility, challenging citizens to navigate a torrent of rapidly evolving, often contradictory information while determining what to believe. This study investigates the cognitive mechanisms underlying epistemic self-efficacy - the perceived ability to distinguish accurate news from misinformation - across different information channels during this high-stakes election cycle. Drawing on data from the Pew Research Center's American Trends Panel (Wave 155, September 2024, N = 9, 360), we test three hypotheses: (H1) whether reliance on social media predicts lower epistemic self-efficacy compared to mainstream news sources; (H2) whether perceived exposure to inaccurate information mediates this relationship; and (H3) whether information fatigue moderates the cognitive burden of verification across platforms. Contrary to expectations rooted in algorithmic filtering theory, we find no significant differences in reported difficulty determining truth between social media and mainstream news users. Instead, epistemic burden is driven by demographics (age, education) and universal information fatigue, suggesting a "leveling" of the information landscape during periods of extreme volatility. This finding challenges platform-deterministic theories and suggests that interventions to support informed citizenship must address cognitive resilience and attention management rather than platform choice alone.

## 1. Introduction

The 2024 U.S. Presidential Election unfolded within an information environment of unprecedented volatility. Occurring against the backdrop of an attempted assassination of a former president and the historic late-stage withdrawal of a sitting president, the election cycle placed extraordinary demands on the American voter. To perform their civic duty, citizens were required not only to consume news but to actively filter a torrent of rapidly evolving, often contradictory information. In this high-stakes ecosystem, the primary challenge for the electorate may no longer be simply accessing information, but navigating the cognitive burden of verifying it.

For over a decade, scholars have warned that the rise of algorithmic social media would fracture the public sphere into "echo chambers" or "epistemic bubbles," potentially degrading the shared reality necessary for democratic discourse (Nguyen, 2020; Pariser, 2011; Sunstein, 2017). Early research highlighted how these platforms could accelerate the spread of fabricated news stories, leading to verifiable deficits in voter knowledge. However, as the digital landscape has matured, the nature of the problem has shifted. The modern information crisis is less about the *absence* of facts and more about the *overabundance* of noise, i.e., a phenomenon Information Science scholars characterize as "information pollution" or "high-entropy" environments. In such a landscape, the critical variable is not just what voters believe, but their **epistemic self-efficacy**: their perceived ability to successfully distinguish between accurate news and misinformation (Khan & Idris, 2019; Kurbanoglu et al., 2006; Pingree, 2011).

While much attention has been paid to the *content* of misinformation, less is known about the *experience* of the user trying to filter it. Does the algorithmic architecture of social media inherently impose a higher "verification tax" on users compared to traditional editorial media? Or has the proliferation of partisan punditry across cable news and talk radio rendered the mainstream environment just as disorienting?

## 1.1. Problem Statement

Despite the growing concern over "fake news," there remains a critical gap in understanding the psychological toll of the modern information environment. Existing literature often treats truth discernment as a performance metric (i.e., *did the user identify the fake headline?*), neglecting the cognitive cost incurred during the process. This "efficacy gap" is significant because low epistemic self-efficacy can lead to information avoidance, cynicism, and civic disengagement (Pingree, 2011; Pinkleton & Austin, 2004; Skovsgaard & Andersen, 2020).

Furthermore, current research often assumes a binary distinction where "Mainstream Media" is safe and "Social Media" is dangerous. This study challenges that assumption by investigating whether the perceived difficulty of determining truth in the 2024 election was truly a function of platform choice, or if it was driven by broader factors such as demographic divides and universal information fatigue. Understanding this distinction is vital for designing interventions: if the problem is the platform, we need better algorithms; if the problem is the user's cognitive load, we need "slower" news.

## 1.2. Research Question

To address this gap, this study analyzes data from Wave 155 of the Pew Research Center's American Trends Panel (Pew Research Center, 2024), fielded in September 2024. It examines the relationship between a voter's primary information channel and their subjective experience of truth discernment.

The study is guided by the following primary research question:

> **RQ1:** *To what extent does reliance on social media versus mainstream news sources predict a voter's epistemic self-efficacy, i.e., the perceived difficulty of determining what is true and what is not, specifically during the 2024 U.S. Presidential Election?*

To fully unpack this relationship, this study further explores two potential mechanisms:

> **RQ2 (Mediation):** *Is the relationship between news source and difficulty mediated by the frequency of exposure to perceived inaccurate news coverage?*
>
> **RQ3 (Moderation):** *Does information fatigue ("news burnout") moderate this relationship, amplifying the cognitive burden for social media users?*

## 1.3. Significance and Contributions

This study advances the field of Information Science and political communication in three critical ways.

First, **theoretically**, it reorients the scholarly debate from *content analysis* (detecting fake news) to *human-information interaction* (the cognitive burden of verification). By measuring "epistemic self-efficacy" rather than just accuracy, this research captures the human cost of the current information ecosystem. It tests the "platform determinism" hypothesis, i.e., the idea that the medium dictates the epistemic outcome, against a more user-centric model that prioritizes cognitive state (fatigue) and demographics.

Second, **empirically**, it provides a snapshot of voter sentiment during a "black swan" election cycle characterized by extreme uncertainty. While previous studies have analyzed stable media environments, this research offers rare insight into how voters navigate truth-seeking during periods of high volatility, testing whether mainstream journalism retains its stabilizing function when reality itself becomes erratic.

Finally, **practically**, the findings hold significant implications for information literacy education. If the difficulty of determining truth is driven by fatigue rather than platform choice, then traditional "fact-checking" interventions may be insufficient. Instead, this suggests a need for a new pedagogical approach focused on "attention management" and "cognitive resilience," equipping citizens to manage the sheer volume of information without succumbing to epistemic exhaustion.

### 1.4. Structure of the Paper

The remainder of this paper is organized as follows. Section 2 reviews the relevant literature on algorithmic echo chambers, information behavior, and the cognitive mechanisms of truth discernment, establishing the theoretical framework for the study. Section 3 describes the methodology, detailing the survey instrument (Pew Research Center's American Trends Panel), variable operationalization, and the statistical procedures used. Section 4 presents the results of the logistic regression and causal mediation analyses. Finally, Section 5 discusses the implications of the findings, interpreting the "null results" regarding platform effects through the lens of information fatigue and demographic divides, and offers recommendations for future research and information literacy interventions.

### 2. Literature Review

The contemporary information ecosystem is characterized by a structural fracture between traditional, editorially mediated news and algorithmically amplified content, a fundamental alteration of the cognitive architecture of journalism. In this high-stakes environment, particularly during the 2024 election cycle, the core challenge for informed citizenship has migrated from securing information access to mastering information discernment. The sheer volume and velocity of content, coupled with platform design optimized for engagement over veracity, impose a significant psychological cost on the user, i.e., a quantifiable cognitive burden. This literature review establishes the theoretical justification for a conceptual model investigating how the primary **Information Source** (Social Media versus Mainstream News) affects the self-perceived ability to determine truth, namely, **Epistemic Self-Efficacy (ESE)** (Khan & Idris, 2019; Kurbanoglu et al., 2006; Pingree, 2011). The model posits that reliance on social media inherently demands higher cognitive resources for verification than mainstream news, resulting in lower ESE.

The underlying premise of this investigation is that the convenience afforded by the algorithmic distribution of news is deceptive; the platform's structural efficiency is achieved by externalizing the high-cost labor of content verification directly onto the user. The conceptual model is systematically broken down across three hypotheses: Hypothesis 1 establishes the direct effect (the Channel Tax); Hypothesis 2 identifies the mechanistic explanation (the Signal-to-Noise Ratio, or Perceived Inaccuracy); and Hypothesis 3 defines the boundary condition (News Fatigue).

## 2.1. Overview of the Conceptual Model

The model utilizes established constructs from communication and social cognitive theory to map the pathway from media consumption habits to self-perceived civic confidence.

Table 1: Key Constructs and Theoretical Foundations

| Construct | Conceptual Definition | Operationalization Strategy | Theoretical Anchor |
|---|---|---|---|
| **Information Source (IV)** | Primary reliance on algorithmic (SM) vs. editorial (Mainstream) gatekeeping. | Dichotomous or relative usage scale of SM (e.g., Facebook, X) vs. established news brands. | Media System Theory; Platform Affordances |
| **Epistemic Self-Efficacy (DV)** | Self-confidence in the ability to discern factual political claims (difficulty determining truth). | Validated ESE scale items (e.g., "I am confident I can tell the true political claims from the false ones"). | Social Cognitive Theory (Pingree, 2011; Stoddard & Chen, 2018) |
| **Perceived Inaccuracy (Mediator)** | Subjective frequency of exposure to low-quality or misleading information. | Frequency scale of encountering fabricated, sensational, or misleading headlines/posts. | Signal-to-Noise Ratio Theory (Brody, 2022) |
| **News Fatigue (Moderator)** | Psychological exhaustion and feeling overwhelmed or "worn out" by political coverage. | Established News Fatigue scale items (e.g., feelings of overload, desire to avoid news). | Information Processing Theory; Cognitive Load (Metag & Gurr, 2023) |

## 2.2. Theoretical Framework: Epistemic Self-Efficacy and the Crisis of Verification

### 2.2.1. ESE in the Digital Age: Self-Confidence in Political Fact Discernment

Epistemic Self-Efficacy is defined as the self-confidence an individual possesses regarding their ability to determine whether factual claims are accurate (Pingree, 2011). Low ESE is structurally equivalent to a heightened, self-perceived "difficulty determining truth." This measure is an adaptation of the broader efficacy literature, designed to reflect the contemporary need for critical discernment within a complex, participatory media culture (e.g., Stoddard & Chen, 2018).

While conventional wisdom suggests that higher ESE should correlate positively with greater political knowledge and literacy (Zhang et al., 2024), the characteristics of the algorithmic environment complicate this relationship. Personalization algorithms, designed to tailor information to the user's preferences, can inadvertently restrict the range of exposure to content, potentially leading to the formation of an inaccurate mental map of reality (Bahg et al., 2025). This limited exposure risks creating an "illusion of competence," where a social media user may subjectively feel informed and confident (initially scoring high on ESE) despite having limited exposure to diverse or verifiable information. The current model addresses this potential inflation by focusing on the self-perceived *difficulty* of the task, ensuring that the metric captures the subjectively experienced cognitive struggle that results from navigating a high-friction information environment. Consequently, a finding of low ESE serves as the subjectively experienced consequence of navigating a system that imposes an excessive cognitive tax on verification.

### 2.2.2. The Consequences of Diminished Efficacy

The erosion of ESE has direct, measurable consequences for civic participation. When citizens feel misinformed or perceive the news environment as confusing and difficult to navigate, they are significantly more likely to experience news fatigue and subsequently engage in active news avoidance (Hasell & Halversen, 2024). News avoidance, defined as deliberately sidestepping information, is a psychological defense mechanism against the persistent strain of media saturation overload (Song et al., 2017; Skovsgaard & Andersen, 2020).
Research has established that the sequence often runs from information overload to decreased news efficacy, which in turn predicts increased news avoidance, particularly on social media (Metag & Gurr, 2023). This behavioral pattern means that citizens who lose confidence in their ability to discern truth not only stop consuming news but also tend to step "out of the conversation completely," including avoiding political discussions with others (Hasell & Halversen, 2024). The failure to maintain ESE thus presents a critical threat to an informed citizenry.

### 2.2.3. Platform Affordances and the Shift in Journalistic Gatekeeping
The decline in Epistemic Self-Efficacy (ESE) must be analyzed in the context of the evolving structure of news delivery. The economic architecture of global information dissemination has fractured, evidenced by the plummeting referral traffic from major social platforms to traditional news publishers (Newman, 2024). This collapse of the 'platform press' model (Rashidian et al., 2020) has exposed the fragility of algorithmic dependence (Nielsen & Ganter, 2018), forcing the industry to pivot from an aggregation model to one based on direct relationships and powered by Artificial Intelligence (Newman, 2024). This migration of value from "distribution" to "cognition" means that algorithms, optimized primarily for engagement, now effectively function as the 'gatekeESErs', prioritizing content based on its potential to stimulate reactions (likes, shares, comments), rather than journalistic norms like accuracy or objectivity (Choi, 2024).
The key implication for the user is the demise of traditional journalistic gatekeeping. In the absence of professional editorial curation, the responsibility for validating the veracity of information, i.e., the ultimate act of cognitive labor, is shifted entirely to the consumer.

## 2.3. Hypothesis 1 Rationale: The Channel Effect and the Zipf Principle of Least Effort
Hypothesis 1 investigates a fundamental structural difference in news environments, focusing on the immediate cognitive cost imposed by the communication channel itself. This analysis is grounded in established principles of human behavior and cognitive economics.

### 2.3.1. Zipf's Principle of Least Effort (PLE): A Structural Constraint on Cognition
Zipf's Principle of Least Effort (PLE) is a robust empirical regularity asserting that human systems, including language and action, seek to minimize effort while maximizing efficiency (Zipf, 1949; Ferrer i Cancho & Solé, 2003). Applied to modern news consumption, the PLE suggests that individuals will gravitate toward the path of least cognitive resistance to satisfy their information needs.
The difference between mainstream and social media news sources can be modeled as a difference in the cost of information verification. A channel that minimizes the user's intrinsic need for verification aligns with the PLE; a channel that maximizes verification labor violates it.

### 2.3.2. Structural Verification Costs: Editorial vs. Algorithmic
Mainstream news organizations, despite their own challenges in the digital era, function under an implicit editorial model. This model leverages institutional credibility, adherence to professional journalistic norms (fairness, accuracy, objectivity; Choi, 2024), and standardized editorial processes. For the consumer, this structure means the verification cost is largely *prepaid* by the institution; the user's cognitive tax is reduced to applying a simple trust heuristic (e.g., "Do I trust this brand?"). The effort required for factual discernment is inherently low.
Conversely, social media platforms are uncurated, algorithmic environments. They operate outside conventional journalistic accountability, and the content flow is saturated with information from disparate sources, often lacking institutional fact-checking overlays. This forces the user into a constant, high-effort verification cycle for every discrete piece of information encountered (Le et al., 2025). The cognitive structure of the social media environment inherently demands higher labor, thereby imposing a significantly higher cognitive tax on verification.

### 2.3.3. The Personalization Trap and the Cognitive Barrier
Social media's mechanism of personalization initially appears to satisfy the Principle of Least Effort. Users benefit from the "News Finds Me" perception, believing they are well-informed without actively seeking political information, because the content comes directly to them (Song et al., 2020). This initial convenience is the structural trap.
However, the continuous nature of social media interaction, which positively associates platform use (such as access through Facebook) with perceived news overload (Lee, Lindsey, & Kim, 2017), ultimately transforms this convenience into cognitive debt. The algorithmic amplification creates an isolated domain of information, which actively hinders the formation of a holistic and accurate understanding. The low ESE hypothesized for social media users is the consequence of this system failing: it represents the moment the structural cost of verification is realized. Users report difficulty determining truth because they are suddenly faced with the high, unpaid cognitive labor that the algorithmic structure had previously obscured. The study therefore tests whether the psychological satisfaction derived from algorithmic convenience outweighs the subsequent cognitive collapse when the user must perform high-stakes political discernment.

> **Hypothesis 1:** *Respondents who rely on social media as their primary news source will report significantly lower epistemic self-efficacy (higher difficulty in determining truth) compared to respondents who rely on mainstream news organizations.*

### 2.4. Hypothesis 2 Rationale: Perceived Inaccuracy as the Signal-to-Noise Mechanism
Hypothesis 2 establishes the mechanism through which the Channel Effect (H1) operates. It utilizes concepts from information science to explain why social media users experience such difficulty: the perceived low quality of information environment fundamentally degrades their confidence in truth discernment.

### 2.4.1. Conceptualizing Misinformation through SNR Theory
The theoretical foundation for the proposed mediation mechanism is the Signal-to-Noise Ratio (SNR), a fundamental metric in Information Theory that quantifies the strength of a desired signal relative to unwanted background interference (Shannon, 1948). While traditionally applied to telecommunications to measure channel capacity, this study adapts the SNR framework to the cognitive processing of political information.
In this context, the "Signal" is defined as verifiable, high-integrity information necessary for informed civic participation. Conversely, "Noise" encompasses all disruptive inputs, specifically inaccurate, irrelevant, or intentionally misleading content that degrades the fidelity of the message. This approach allows for a unified mathematical and conceptual representation of disparate information types, treating reliable news as "signal" and misinformation as "noise" (Brody, 2022).

### 2.4.2. The Mediation Path: Low SNR Leads to Low ESE

The mediation path hypothesized in H2 connects the informational environment to the psychological outcome.

**Source → Perceived Inaccuracy (Low SNR):**
Reliance on social media is predicted to correlate with a subjectively higher frequency of exposure to inaccurate information. This is directly attributable to algorithmic platform affordances, which amplify sensational material based on engagement metrics rather than editorial vetting (Napoli, 2019; Vosoughi et al., 2018). Sophisticated hoaxes, such as the AI-generated fabricated image of the Pentagon explosion in 2023 that temporarily disrupted the stock market, illustrate the functional reality of a low SNR environment (Associated Press, 2023). In this ecosystem, false claims can trend rapidly and credibly, exploiting algorithmic trading systems that prioritize speed over verification (Di Minin & Nasi, 2023). The user's perception of inaccuracy directly reflects the objective failure of the platform to maintain a high SNR.

**Perceived Inaccuracy → Low ESE (Cognitive Overload):**
When users subjectively perceive that the volume of noise (inaccuracy and irrelevance) is overwhelming, the effort required to extract the reliable signal becomes cognitively taxing. This experience of overwhelming complexity contributes to the decline in ESE, manifesting as the belief that the task of discernment is excessively burdensome or futile (Hasell & Halversen, 2024).

It is important to recognize that the perceived inadequacy of the information environment is a composite psychological mediator. While traditional SNR focuses on separating signal from noise (falsehoods), social media also introduces significant amounts of *information irrelevance* (personal updates, non-news clickbait) that contribute to information fatigue (Le et al., 2025). This requires the user to perform a dual-filtering task: first, filtering for relevant news content, and second, filtering that content for veracity. This compounded cognitive demand accelerates the sensation that the news environment is unnavigable, leading to lowered self-confidence in the ability to process it successfully.

### 2.4.3. Scholarly Validation of Perceived Quality

The mediating role of perceived quality is essential to understanding information processing outcomes (Pelau et al., 2023). Americans maintain high expectations for news to be fair, accurate, and objective (Choi, 2024). The observation that platform reliance leads to the perception of frequent inaccuracy confirms that the psychological toll of social media news is a function of the content quality failing to meet these civic expectations. This failure, quantified through the low SNR, directly explains the reported loss of self-confidence in political discernment.

> **Hypothesis 2:** *The relationship between news source type and verification difficulty is mediated by perceived information quality. Specifically, reliance on social media predicts higher frequency of exposure to inaccurate information, which in turn predicts higher difficulty in discerning truth.*

## 2.5. Hypothesis 3 Rationale: Information Fatigue as a Boundary Condition for Cognitive Processing

Hypothesis 3 introduces a critical psychological variable, News Fatigue, as a moderator. It seeks to understand the boundary conditions under which the cognitive tax of social media becomes unbearable, amplifying the negative effect on ESE.

### 2.5.1. The Psychological Reality of News Overload

News Fatigue is a recognized psychological state defined as exhaustion resulting from the incessant and overwhelming flow of news content across both traditional and social media platforms (Song et al., 2017). While not a new phenomenon, its prevalence has significantly increased over the past decade due to the internet-driven explosion in content, which eliminated the previous temporal and volume limits imposed by traditional media (Schmitt et al., 2018). This psychological depletion is widespread, with sixty-five percent of U.S. adults reporting the need to actively limit political media consumption due to fatigue (AP-NORC, 2024).
The state of fatigue represents cognitive resource depletion. It impairs the user's ability to process, filter, and critically evaluate incoming information effectively (Lang, 2000; Sweller, 2011). The empirical association between issue fatigue and negative evaluations of media coverage quality, including perceived lack of credibility, confirms that depletion biases perception negatively across the board (Gurr & Metag, 2021).

### 2.5.2. The Moderating Effect of Depletion

Fatigue functions as a boundary condition (moderator) for the source-efficacy relationship. High News Fatigue universally depletes cognitive reserves, making any effortful task, including verification, harder. Consequently, respondents reporting high fatigue are expected to report generally lower ESE regardless of their primary news source.
However, the hypothesis posits a significant amplified interaction effect. The user relying on social media is already situated in a high-demand environment (H1), where structural verification effort is maximized. When this high-demand channel is accessed by a user whose cognitive resources are already depleted (high fatigue), the combined stressor accelerates the loss of efficacy. The exhausted social media user faces an insurmountable task, as access through platforms like Facebook is empirically associated with high perceived overload (Lee, Lindsey, & Kim, 2017).
This compounded interaction drives the negative ESE to its maximum predicted extent. The individual is not merely tired; they are trying to perform a high-effort task (verification) within a high-noise environment (inaccuracy) while operating with severely depleted cognitive reserves (fatigue).

### 2.5.3. Fatigue as a Factor in Disparate Political Engagement

Understanding this moderating effect is crucial for explaining patterns of civic disengagement in the 2024 election cycle. The moderator provides a critical psychological variable explaining differential news avoidance. Research indicates that when people feel difficulty in evaluating information online, they actively avoid news. Furthermore, these behavioral responses are often conditional on political affiliation: one study found that when feeling misinformed, strong Republicans reported using less news media overall, while strong Democrats relied more on nonpartisan media (Hasell & Halversen, 2024).

If News Fatigue is found to amplify the negative impact of social media reliance on ESE, it implies that the feeling of being "worn out" acts as a non-partisan engine of disengagement, accelerating the loss of confidence that subsequently leads to partisan or non-partisan news avoidance. This mechanism underscores that efforts to maintain an informed electorate must address not only the quality of information but also the psychological capacity of the recipient to process it.

> **Hypothesis 3:** *News Fatigue moderates the relationship between source and efficacy; respondents who report being "worn out" by coverage will report lower truth discernment efficacy across all platforms, but the effect will be most pronounced among social media users due to the high volume of uncurated content.*

## 2.6. Implications for Mitigation and Policy

The results of this investigation offer direct pathways for informed intervention.

1. **Structural Interventions (H1):** If the Channel Effect is empirically dominant, it suggests that merely educating the user is insufficient; structural changes to platform design that reduce the inherent cognitive tax on verification are necessary.
2. **Content Interventions (H2):** A strong mediation effect via Perceived Inaccuracy underscores the need to improve the SNR. This involves increasing the visibility and authority of credible signals, making fact-checking interventions more accessible, and utilizing technology to identify dubious claims.
3. **Psychological Interventions (H3):** If the moderation effect is significant, educational efforts must incorporate strategies for managing media overload and establishing consumption limits, recognizing that the user's psychological state directly dictates their effectiveness as an information processor.

Ultimately, this research confirms that confidence in finding political truth (ESE) is a key element in successfully confronting and discerning misinformation. By quantifying the cognitive burdens imposed by current platform designs, the study provides necessary data to support efforts aimed at bolstering civic confidence rather than simply focusing on media literacy skills in isolation.

## 3. Methodology

### 3.1. Data Source and Sample

This study utilizes data from Wave 155 of the American Trends Panel (ATP), conducted by the Pew Research Center (2024). The survey was fielded from September 16 to September 22, 2024. The ATP is a nationally representative panel of U.S. adults. The survey was administered via a mixed-mode format (Web and CATI/Phone), though the majority of questions regarding digital media consumption utilized web-specific logic.

For the purposes of this analysis, respondents who refused to answer (coded as 99) or responded "Don't know" (coded as 98) on the primary variables of interest were treated as missing data and excluded listwise.

## 3.2. Measures
### 3.2.1. Independent Variable: Primary Information Channel
To test the "Channel Effect" (H1), we operationalized the respondent's primary information environment using the item ELECT_MOST_W155. Participants were asked, "What is the most common way you get political and election news?".
Responses were collapsed into a categorical variable with three levels:
- **Social Media:** Respondents who selected "Social media".
- **Mainstream Media:** A composite group including "Print newspapers or magazines", "News websites or apps", "Radio", and "Television".
- Other/Search: Including "Search through Google", "Podcasts", and "Some other way". Mainstream Media served as the reference group for all regression analyses.

### 3.2.2. Dependent Variable: Epistemic Self-Efficacy
The outcome variable measures the cognitive burden of verification. It is derived from the item CAMPINFODIFF_W155, which asked: "When you get news and information about the presidential campaign and candidates, do you generally find it...".
The variable was recoded into a binary metric of **Low Efficacy/High Burden**:
- **0 (High Efficacy):** "Easy to determine what is true and what is not".
- **1 (Low Efficacy):** "Difficult to determine what is true and what is not".

### 3.2.3. Mediating Variable: Perceived Signal-to-Noise Ratio
To test the mechanism of "Information Pollution" (H2), we utilized the item INACCMED_W155. Participants were asked: "How often have you seen inaccurate news coverage about the 2024 presidential election?"
The original 5-point scale ranged from "Extremely often" to "Not at all". This was reverse-coded so that higher values indicate a higher frequency of perceived noise (1 = Low Noise, 5 = High Noise/Extremely Often).

### 3.2.4. Moderating Variable: News Fatigue
Information overload was measured using ELECTFTIGUE_W155. Respondents were asked to choose which statement came closer to their view: "I like seeing a lot of coverage..." or "I am worn out by so much coverage of the campaign and candidates". This was coded as a binary variable (0 = High Interest, 1 = High Fatigue).

## 3.3. Analytical Strategy
All statistical analyses were conducted using the R statistical computing environment.
To test **Hypothesis 1** (The Channel Effect), a binomial logistic regression was employed to predict the likelihood of Low Epistemic Self-Efficacy based on the Primary Information Channel, controlling for standard demographics (age, education).
To test **Hypothesis 2** (Signal-to-Noise Mechanism), a causal mediation analysis was conducted using the mediation package in R. We estimated the Average Causal Mediation Effect (ACME) to determine if the impact of Social Media usage on Epistemic Self-Efficacy is transmitted through the mediator of Perceived Inaccuracy (INACCMED_W155).
To test **Hypothesis 3** (The Moderation of Fatigue), an interaction term (Primary Channel × News Fatigue) was added to the logistic regression model to assess if information fatigue significantly amplifies the cognitive burden for social media users specifically

## 4. Analysis

The analysis proceeds in two distinct phases: first, an **Exploratory Data Analysis (EDA)** to assess the distributional properties of the core variables and verify the assumptions of the proposed models; and second, **Inferential Hypothesis Testing** to evaluate the relationships between media source, perceived noise, and epistemic self-efficacy. Consistent with the data cleaning procedures outlined in the methodology, listwise deletion was applied to handle missing data for the variables of interest, resulting in a final analytical sample of 9,360. The results for each hypothesis are reported sequentially below, accompanied by relevant statistical tables and visualizations.

### 4.1. Exploratory Data Analysis

Prior to formal hypothesis testing, a comprehensive exploratory data analysis (EDA) was conducted in R to assess the distributional properties of key variables and verify statistical assumptions.

**Distributional Checks**: Univariate analysis was performed on the primary independent variable, ELECT_MOST_W155, to ensure sufficient sample size within the "Social Media" subgroup for reliable logistic regression (Figure 1). We also examined the frequency distribution of the outcome variable, CAMPINFODIFF_W155, to verify that the binary classes (High vs. Low Efficacy) were not severely imbalanced (Figure 2), which could bias model probability estimates.

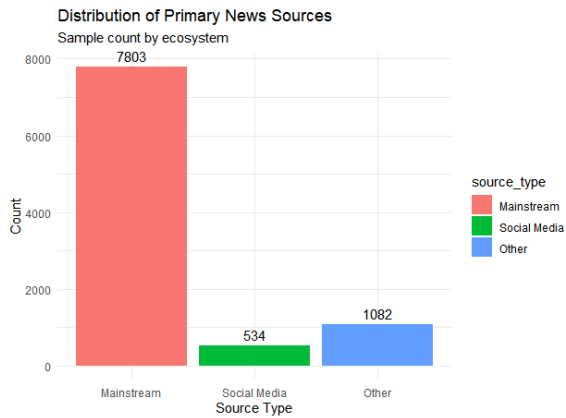
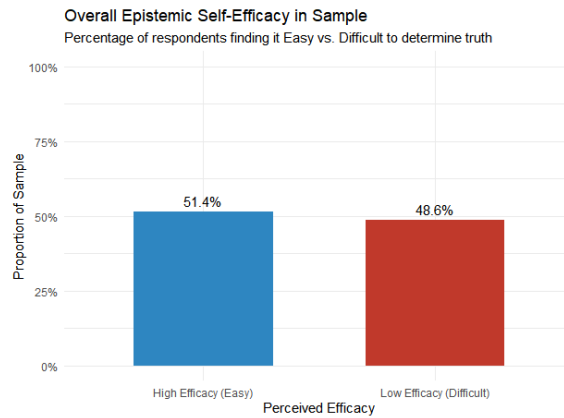

*Figure 1 Distribution of Primary News Sources*     *Figure 2 Overall Epistemic Self-Efficacy in Sample*

**Bivariate and Interaction Visualizations**: To preliminarily assess the viability of the hypothesized relationships, we generated visual summaries:
- Source vs. Efficacy (H1): Grouped bar charts were inspected to compare the raw proportions of respondents reporting low efficacy across the three information source categories (Mainstream, Social Media, Other, Figure 3).

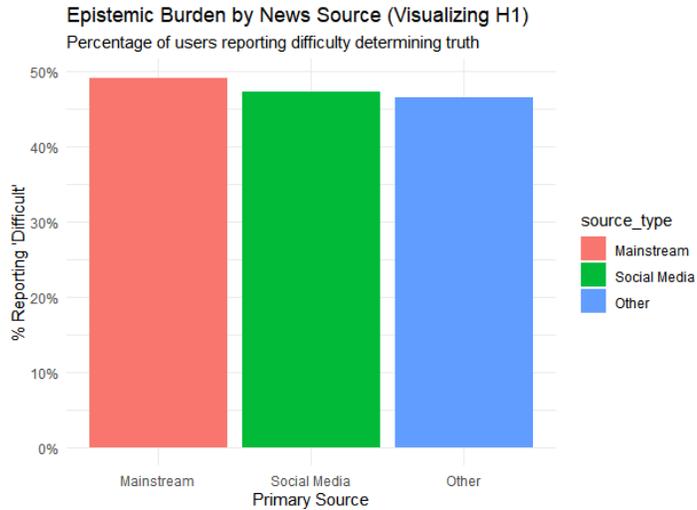
*Figure 3 Epistemic burden by News Source*

- Source vs. Noise (H2): Density plots were generated to visualize the distribution range of Perceived Information Noise (INACCMED_W155) across source types. This step allowed for the detection of potential differences among the source types and confirmed the ordinal nature of the noise perception scale (Figure 4).

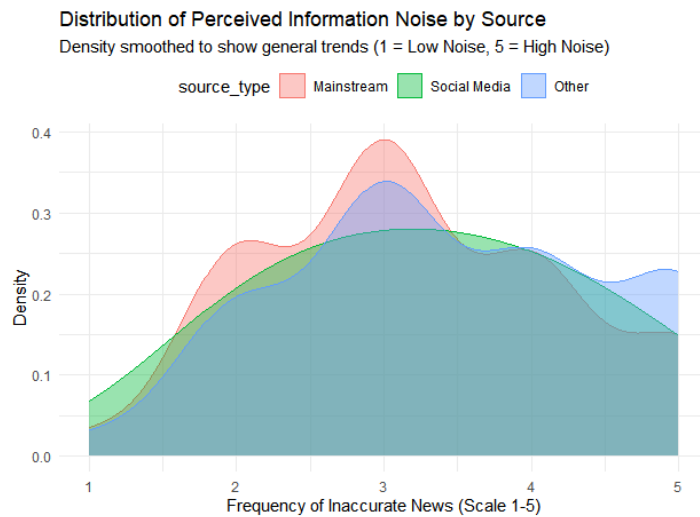
*Figure 4 Distribution of Perceived Information Noise by Source*

- Fatigue Interaction (H3): Interaction plots were created to observe the slope differences in efficacy between high-fatigue and low-fatigue groups within each media ecosystem, helping to visualize the potential moderating effect of ELECTFTIGUE_W155 before specifying the interaction term in the model (Figure 5).

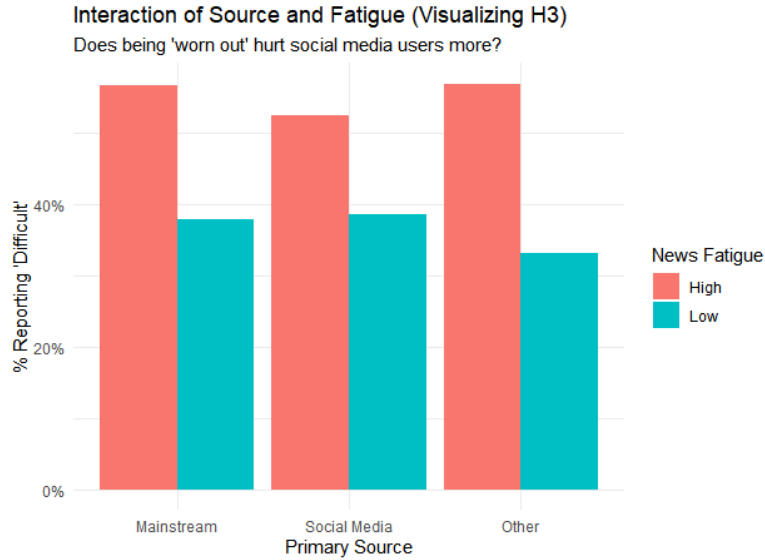
*Figure 5 Interaction of Source and Fatigue*

**Missing Data Analysis:** We analyzed the patterns of non-response (responses coded as 98 "Don't Know" or 99 "Refused") across all four core variables. Cases with missing data on the dependent variable or primary predictor were assessed to determine if they occurred at random or were correlated with specific demographic profiles. Based on this assessment, a listwise deletion strategy was deemed appropriate for the final analytical models.

**4.2. The Effect of Information Channel on Epistemic Efficacy (H1)**
To test Hypothesis 1, which posited that reliance on social media leads to lower epistemic self-efficacy compared to mainstream media, a binomial logistic regression was conducted. The model predicted the likelihood of respondents reporting difficulty in determining truth based on their primary news source, controlling for education and age.

The results, presented in **Table 1**, indicate that Hypothesis 1 was **not supported**. Contrary to the expectation that algorithmic environments impose a higher verification burden, the primary news source was not a significant predictor of efficacy. Respondents who rely on social media *(β = -0.03, SE = 0.09, p = .727)* and "Other" sources *(β = -0.09, SE = 0.07, p = .159)* showed no statistically distinguishable difference in reported difficulty compared to mainstream media consumers.

However, the analysis revealed significant disparities based on demographic control variables, suggesting an "epistemic divide" rooted in age and educational attainment rather than platform choice:

- **Education:** Higher educational attainment was strongly associated with higher self-efficacy. Compared to those with less than a high school education, college graduates *(β = -0.27, p < .05)* and postgraduates *(β = -0.62, p < .001)* were significantly less likely to report difficulty in discerning the truth.
- **Age:** Younger adults (18-29) appear to experience the highest cognitive burden. Compared to this reference group, respondents aged 50-64 *(β = -0.26, p < .001)* and those over 65 *(β = -0.32, p < .001)* were significantly less likely to find the information environment difficult to navigate.

These findings suggest that the struggle to discern truth in the 2024 election is a generalized phenomenon affecting users across all media ecosystems, with the burden falling disproportionately on younger and less educated voters, regardless of whether they consume news via TikTok or cable television.

*Table 1: The result of the statistical analysis for H1*

| Predictor | B | SE | Odds Ratio (eB) | p-value |
|---|---|---|---|---|
| (Intercept) | 0.39 | 0.13 | 1.48 | .003** |
| **Information Channel** | | | | |
| Mainstream Media (Ref) | — | — | — | — |
| Social Media | -0.03 | 0.09 | 0.97 | .727 |
| Other Sources | -0.09 | 0.07 | 0.91 | .159 |
| **Education Level** | | | | |
| Less than High School (Ref) | — | — | — | — |
| High School Grad | 0.03 | 0.13 | 1.03 | .830 |
| Some College | -0.09 | 0.13 | 0.91 | .473 |
| Associate Degree | -0.16 | 0.13 | 0.86 | .245 |
| College Graduate | -0.27 | 0.12 | 0.76 | .028* |
| Postgraduate | -0.62 | 0.13 | 0.54 | <.001*** |
| **Age Group** | | | | |
| 18–29 (Ref) | — | — | — | — |
| 30–49 | -0.09 | 0.08 | 0.92 | .254 |
| 50–64 | -0.26 | 0.08 | 0.77 | <.001*** |
| 65+ | -0.32 | 0.08 | 0.72 | <.001*** |

Note: N = 9,367. AIC = 12,846.
Dependant Variable: Low Efficacy (1 = Difficult to determine truth, 0 = Easy).
Odds Ratios (OR) < 1 indicate a lower likelihood of reporting difficulty compared to the reference group.
\* $p < .05$, \*\* $p < .01$, \*\*\* $p < .001$.

## 4.3. Mediation Analysis of Perceived Information Noise (H2)

Hypothesis 2 proposed that the relationship between information source and epistemic self-efficacy is mediated by the perceived "signal-to-noise" ratio, specifically, that social media users would encounter higher frequencies of inaccurate information (INACCMED_W155), thereby increasing the cognitive burden of verification.

To test this mechanism, a causal mediation analysis was conducted using nonparametric bootstrapping with 1,000 simulations. The results are summarized in **Table 2**.

The analysis revealed that the Average Causal Mediation Effect (ACME) was not statistically significant (Estimate = 0.005, 95% CI [-0.001, 0.012], $p = .118$). This indicates that perceived exposure to inaccurate news does not significantly mediate the relationship between platform choice and the difficulty of discerning truth. Furthermore, the Average Direct Effect (ADE) remained insignificant (Estimate = -0.013, $p = .550$).

Consequently, **Hypothesis 2 is not supported.** The data does not provide evidence that the "information pollution" mechanism creates a unique epistemic disadvantage for social media users relative to mainstream news consumers in this election cycle.

*Table 2: Causal Mediation Analysis Results (Source -> Noise -> Efficacy)*

| Effect | Estimate | 95% CI Lower | 95% CI Upper | p-value |
|---|---|---|---|---|
| ACME (Indirect Effect) | 0.005 | -0.001 | 0.012 | .118 |
| ADE (Direct Effect) | -0.013 | -0.060 | 0.031 | .550 |
| Total Effect | -0.008 | -0.055 | 0.038 | .738 |

**Note:** N = 9,364. Simulations = 1,000. ACME = Average Causal Mediation Effect. ADE = Average Direct Effect. Confidence intervals calculated using the percentile method.

### 4.4. The Moderating Role of Information Fatigue (H3)

Finally, Hypothesis 3 posited that information fatigue would moderate the relationship between source type and epistemic self-efficacy, specifically hypothesizing that the cognitive burden of social media would be amplified among users who reported being "worn out" by election coverage.

To test this, an interaction term (Source Type × Fatigue) was added to the logistic regression model. The results, detailed in **Table 3**, indicate that **Hypothesis 3 is not supported**. The interaction between relying on social media and reporting low fatigue was not statistically significant ($\beta = 0.20, p = .277$). This suggests that the relationship between fatigue and difficulty determining truth is consistent across media ecosystems; social media users are not uniquely susceptible to the deleterious effects of information overload compared to mainstream news consumers.

However, the analysis identified a powerful **main effect of fatigue** ($\beta = -0.75, p < .001$). Regardless of their primary news source, respondents who reported being "worn out" by the coverage were significantly more likely to report difficulty in discerning the truth compared to those who enjoyed the coverage. Along with the demographic predictors identified in H1, information fatigue appears to be a primary driver of epistemic burden in the 2024 election cycle, functioning as a universal stressor rather than a platform-specific phenomenon.

*Table 3: Logistic Regression Predicting Low Efficacy with Fatigue Interaction*

| Predictor | B | SE | z | p-value |
|---|---|---|---|---|
| (Intercept) | 0.74 | 0.14 | 5.41 | <.001*** |
| **Source Type (Ref: Mainstream)** | | | | |
| Social Media | -0.13 | 0.12 | -1.10 | .271 |
| Other | 0.02 | 0.09 | 0.26 | .797 |
| **Fatigue (Ref: High/Worn Out)** | | | | |
| Low Fatigue (Likes Coverage) | -0.75 | 0.05 | -15.83 | <.001*** |
| **Interactions** | | | | |
| Social Media \times Low Fatigue | 0.20 | 0.19 | 1.09 | .277 |
| Other \times Low Fatigue | -0.21 | 0.14 | -1.55 | .120 |
| **Controls** | | | | |
| Education (Postgrad vs. <HS) | -0.67 | 0.13 | -5.27 | <.001*** |
| Age (65+ vs. 18–29) | -0.28 | 0.08 | -3.53 | <.001*** |

**Note**: N = 9,364. AIC = 12,529.
Dependant Variable: Low Efficacy (1 = Difficult to determine truth).
Reference group for Fatigue is "High Fatigue" (Worn out).
\* p < .05, \*\* p < .01, \*\*\* p < .001.

## 5. Discussion & Conclusion

This study sought to determine whether the "epistemic crisis" of the 2024 U.S. Presidential Election was primarily a function of where voters consumed information. Drawing on theories of algorithmic filtering and echo chambers, we hypothesized that reliance on social media would uniquely predict greater difficulty in distinguishing truth from falsehood (H1), mediated by higher exposure to information "noise" (H2), and exacerbated by news fatigue (H3).
Contrary to these expectations, the data revealed a striking pattern of "epistemic equivalence." We found no statistically significant difference in epistemic self-efficacy between social media users and mainstream news consumers. Instead, the struggle to verify truth was driven almost exclusively by **who the user is** (demographics) and **their cognitive state** (fatigue), rather than the technological architecture of their news feed.

## 5.1. The "Leveling" of the Information Landscape

The rejection of Hypothesis 1 and 2 challenges the prevailing academic narrative that social media is a uniquely degraded information environment compared to traditional journalism. In this sample, mainstream media consumers were just as likely to report being overwhelmed and confused as those scrolling through algorithmic feeds.

This counter-intuitive finding may be explained by the specific peculiarities of the 2024 election cycle. The period covered by the survey (September 2024) followed a cascade of high-complexity, high-volatility events, including the attempted assassination of Donald Trump and the unprecedented withdrawal of President Joe Biden from the race. These "black swan" events may have created a **saturation of uncertainty** that transcended platform boundaries. When reality itself becomes volatile, mainstream journalism, which often relies on speculation and pundits during breaking news, may offer no greater clarity than social media commentary, effectively "leveling" the difficulty of verification across all ecosystems.

From an Information Science perspective, this suggests a shift from "platform-specific pollution" to **"ubiquitous ambiguity."** The "signal" in 2024 was so erratic that no channel provided a clear advantage in decoding it.

## 5.2. Fatigue as the Universal Antagonist

While platform choice failed to predict epistemic burden, **Information Fatigue** emerged as a powerful, universal predictor (H3 main effect). Respondents who reported being "worn out" by coverage were significantly more likely to find truth discernment difficult ($p < .001$), regardless of their news diet.

This aligns with the **Cognitive Load Theory**, suggesting that the primary barrier to informed citizenship in 2024 was not "fake news" per se, but the sheer *volume* and *velocity* of information. When users reach cognitive exhaustion, their critical faculties degrade, making verification feel impossible. The fact that this relationship did not interact with source type implies that "Mainstream Fatigue" is just as cognitively taxing as "Algorithmic Fatigue." The 24-hour cable news cycle, with its repetitive "breaking news" banners, appears to deplete user efficacy just as effectively as an infinite scroll.

## 5.3. The Paradox of the Digital Native

Perhaps the most significant finding is the demographic inversion of epistemic efficacy. Younger adults (18-29), traditionally viewed as "digital natives", reported significantly *higher* difficulty in determining truth compared to older cohorts (50+).

This challenges the assumption that digital literacy confers an advantage in online information seeking. Instead, it supports the **"Paradox of Choice"** framework: younger users, who often navigate a more fragmented and multi-stream media diet, may be suffering from "choice overload" in verification. Older adults, who may rely on fewer, established heuristics (e.g., "I trust Cronkite-style delivery"), report higher efficacy not necessarily because they are better at fact-checking, but because their information habits are more bounded and less cognitively demanding.

### 5.4. Limitations
Several limitations temper these conclusions. First, the sample of primary social media users was relatively small (5.7%) compared to mainstream consumers, potentially limiting the statistical power to detect subtle platform-specific effects. Second, the variable "Social Media" collapses distinct architectures (e.g., TikTok's "For You" page vs. Facebook's "News Feed") into a single category, masking potential differences between text-based and video-based algorithms. Finally, self-reported "difficulty" is a measure of perception, not performance; older adults may report "ease" simply because they are more overconfident in false beliefs, a phenomenon known as the Dunning-Kruger effect.

### 5.5. Conclusion and Future Directions
The 2024 election data suggests that the "Epistemic Crisis" is no longer a problem of *access* to high-quality sources, but a problem of **human cognitive limits**.

Future research should pivot away from comparing "Newspapers vs. Facebook" and instead investigate **"Resilience vs. Exhaustion."** Specifically, scholars should examine:
1. **Platform Granularity:** Does the cognitive burden differ between "lean-back" platforms (TikTok/TV) and "lean-forward" platforms (Search/Reddit)?
2. **Intervention Testing:** Can "slow news" formats or algorithmic "friction" reduce fatigue and restore epistemic self-efficacy?
3. **Longitudinal Fatigue:** Does the "worn out" effect accumulate over the election cycle, and at what point does it trigger total news avoidance?

Ultimately, this study indicates that fixing the algorithm may not be enough if the user is too exhausted to process the truth.